\pdfoutput=1

\documentclass[twocolumn,
amsmath,amssymb,
aps,prl,reprint, nofootinbib, nobibnotes,
superscriptaddress
]{revtex4-1}

\usepackage{graphicx}
\usepackage{dcolumn}
\usepackage{float}
\usepackage{bm}
\usepackage{graphicx}
\usepackage[toc,page]{appendix}
\usepackage{blindtext}
\usepackage{physics}

\usepackage{pgfplots}
\pgfplotsset{compat=1.13}

\usepackage[version=4]{mhchem}

\usepackage{hyperref}

\usepackage[overlay,absolute]{textpos}
\newcommand\PlaceText[3]{
	\begin{textblock*}{10in}(#1,#2)  
		#3
	\end{textblock*}
}

\begin{document}
	
\title{Entangling an arbitrary pair of qubits in a long ion crystal}

\author{Pak Hong Leung}
\email{james.leung@duke.edu}
\affiliation{Department of Physics, Duke University, Durham, North Carolina 27708, USA}
\author{Kenneth R. Brown}
\affiliation{Department of Electrical and Computer Engineering, Duke University, Durham, North Carolina 27708, USA}

\date{\today}

\begin{abstract}

 	It is well established that the collective motion of ion crystals can be used as a quantum bus for multi-qubit entanglement. However, as the number of ions increases, it becomes difficult to directly entangle ions far apart and resolve all motional modes of the ion crystal. We introduce a scalable and flexible scheme for efficient entanglement between any pair of ions within a large ion chain, using an evenly distributed 50-ion crystal as an example. By performing amplitude and frequency modulation, we find high-fidelity pulse sequences that primarily drive a transverse motional mode with a wavelength of 4 ion spacings. We present two $500 \mu s$ pulses that can in theory suppress gate errors due to residual motion to below $10^{-4}$, and observe a trade-off between gate power and robustness against unwanted frequency offsets.
 	
\end{abstract}

\maketitle

One important challenge for the quantum information community is to scale up the number of controllable qubits. An exciting motivation would be to solve a computational problem or to simulate a physical system beyond the reach of classical computers. We are approaching the so-called Noisy Intermediate-Scale Quantum (NISQ) regime \cite{nisq}, where gate error is the limiting factor for the width and depth of a quantum circuit. It is estimated that 50 qubits require more memory to simulate than what modern supercomputers can offer \cite{qs1}, and coherent control with 90 qubits may be sufficient to demonstrate quantum supremacy \cite{qs2}. Moreover, large quantities of qubits are required to implement certain quantum error correction schemes with high fault-tolerant thresholds (physical error rate $< 10^{-2}$), where each logical qubit typically consists of more than 10 physical qubits \cite{muyuan, colin, yu}.

In the past few decades, ion trap experiments have been realized with increasing precision \cite{yb1, single1, single2, optimal_control}, with or without individual qubit addressing, offering various applications in quantum information. With as few as 5 ions, many groups have made proof-of-principle demonstrations of simple algorithms \cite{small_computer}, error correction \cite{error_detection}, and quantum correlations \cite{renyi}. With larger numbers of ions, applying a global driving force to the trap allows us to demonstrate large-scale entanglement and quantum phase transitions \cite{14-qubit, 53-qubit}.

Multi-qubit gates remain the limiting factor in terms of both gate time and fidelity. For trapped ions, 2-qubit gates can be mediated through M\o lmer-S\o rensen interaction \cite{Molmer, Sorensen}, where a state-dependent force is applied to trigger the collective motion of an ion lattice. Experimental groups have achieved 2-qubit gate fidelities of higher than 99.99\% for 2-ion traps and about 99\% for 5-ion traps \cite{Lucas1, HF_Wineland}, with gate times 1 to 2 orders of magnitude longer than single-qubit gates. The interaction strength between qubits decays as $1/d^n$ where $n = 3$ in the limit of far detuning from motional sidebands, and $n \approx 1$ or less for near resonance with the common mode \cite{53-qubit}. Thus, the locality of inter-qubit interaction inhibits long-distance controlled operations, which is deemed necessary for the realization of intermediate-scale quantum computers.

In this article, guided by previous experimental success with 5 $\ce{^{171}Yb+}$ ions \cite{my_paper}, we show that it is theoretically possible to entangle any pair of ions in a 50-ion trap with high fidelity using pulse modulation. First, we derive the effective trapping potential where all 50 ions align in one axis with uniform density. We then predict the transverse motional modes of the ion chain, and choose one of them as the main quantum bus for entanglement. For the driving force, we apply a smooth intensity profile along with small frequency oscillations to minimize motional displacement of all modes. Our results show that the total gate error due to residual motion is less than $10^{-4}$, and can tolerate small drifts in trap frequency.

\begin{figure}
	\scalebox{0.3}{\includegraphics{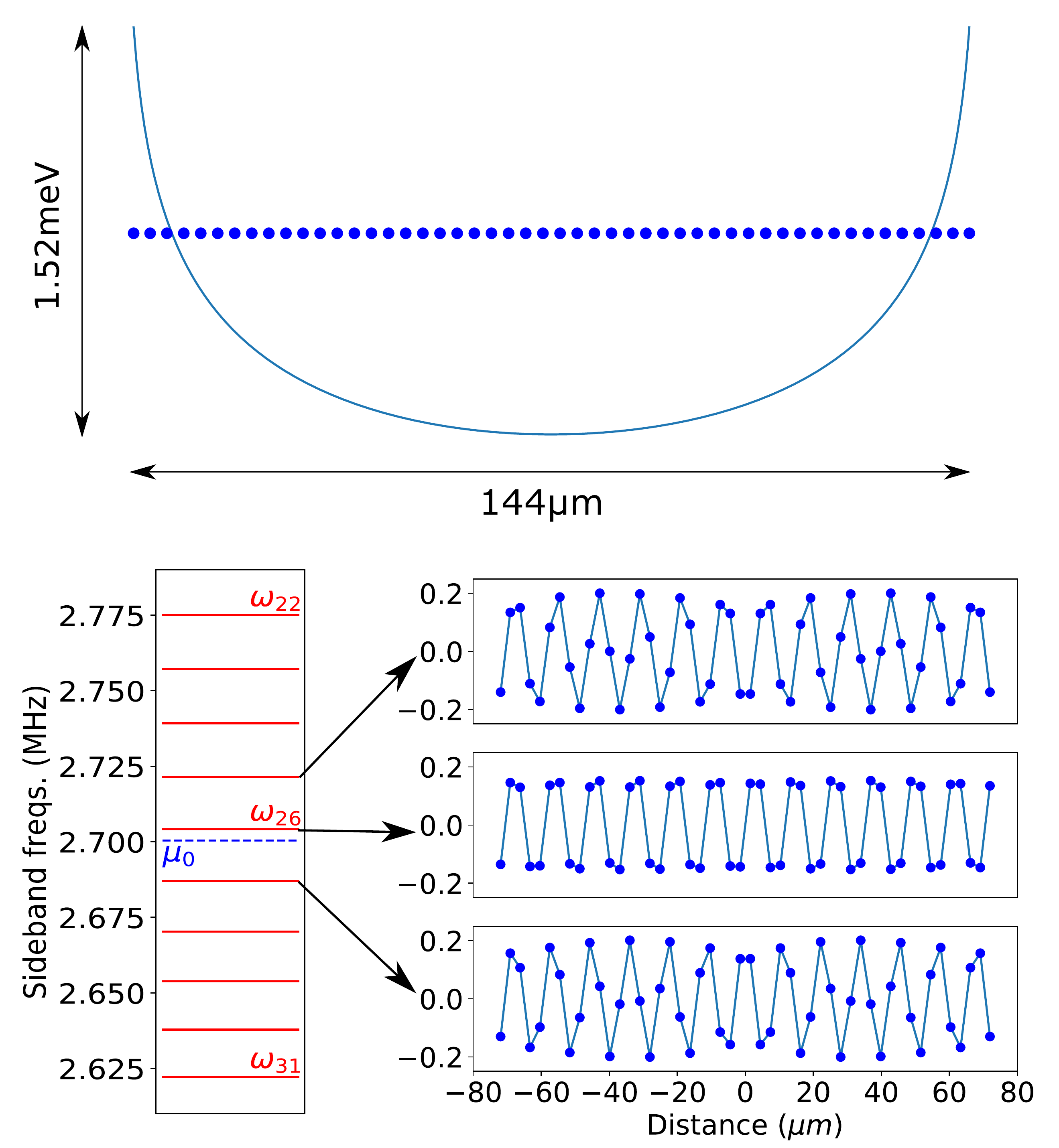}}
	\caption{\small Depiction of how we may choose particular motional modes as the quantum channel for entanglement. (a) Ideal trap potential from equation (2) for r = 0.95 and the corresponding distribution of ions found by gradient descent of electric potential energy. A minimum trap depth of 1.52 meV is required to trap all 50 ions. (b) The middle section of the transverse motional frequencies, showing $\omega_{22}$ to $\omega_{31}$ (solid red lines) and the approximate driving frequency $\mu_0 = \omega_{26} - 3.7$ kHz (blue dashed line), for radial trapping frequency $\omega_x$ = 3.07 MHz and average ion separation = $2.9\mu m$. (c) The 25th, 26th and 27th transverse motional modes (normalized). The uniformity of the 26th mode makes it useful as a channel for arbitrary 2-qubit entanglement.}
\end{figure}

\section{Transverse Motional Modes of an Evenly Spaced Ion Chain}

We begin by outlining the requirements for a suitable trap potential. For a typical 1D ion lattice, radial confinement (x and y-axis) is assumed to be uniform, harmonic, and much stronger than axial confinement (z-axis). In order to trap large numbers of ions ($N \gg 10$), highly anharmonic terms are required in the axial potential. We also need to ensure that all ions are sufficiently separated from their neighbors such that they can be addressed individually using tightly focused laser beams. Previous work with anharmonic traps has used a parameterized form for the potential, then sought to minimize the variance in ion separation \cite{anharmonic1, anharmonic2}. Here we opt for a different approach: we first assume a continuous ion distribution, then integrate to find the potential required to generate it.

We investigate the idealistic case of having $N = 50$ evenly spaced ions across the whole chain, with an average separation $\Delta z \approx 3 \mu$m. In the continuous limit the chain can be modeled as a uniform charge density $\rho_0 = q/\Delta z$. We find an analytical expression for the effective electric field acting on each ion due to Coulombic repulsion:
\begin{equation}
\begin{aligned}
E_{rep}(z) &= \Bigg(\int_{-L}^{z-\epsilon}-\int_{z+\epsilon}^{L}\Bigg)\frac{k\rho_0 dz'}{(z'-z)^2} \\
&= k\rho_0\Bigg(\frac{1}{L-z}-\frac{1}{L+z}\Bigg)
\end{aligned}
\end{equation}
where L is the half length of the ion chain, $\epsilon$ is half the average ion-to-ion separation, and $\rho_0 = q/\Delta z$ is the linear charge density. We integrate again to find the effective trap potential required to maintain the shape of the lattice
\begin{equation}
V_{trap}(z) = rk\rho_0 \text{ln}\Bigg(\frac{L^2}{L^2-z^2}\Bigg)
\end{equation}
where $r \approx 1$ is an additional scaling factor for adjustment (blue curve in Fig. 1(a)). 

To avoid infinite potential walls, we let this expression be valid only for $|z| < sL$ where $s$ is slightly smaller than 1 such that there is a finite upper bound for the electric field, $E_{max}$. The minimum for $E_{max}$ is the electric field experienced by the ions at the edges of the chain, given approximately by 
\begin{equation}
E_{edge} = \sum_{n=1}^{N}\frac{kq}{(n\Delta z)^2} \approx \frac{\pi^2}{6}\frac{kq}{\Delta z^2} \approx 260 \text{V/m}
\end{equation}
for an average separation $\Delta z =$ 3 $\mu$m. This should be well within the maximum field $\sim$ 10 V/100 $\mu$m = $10^5$ V/m that can be generated by typical microfabricated ion traps. Tight laser beamwidths of about 1.5 $\mu$m have been realized in past experiments \cite{small_computer}, allowing us to address any ion with very small crosstalk errors with precision beam steering. \\

\PlaceText{18mm}{20mm}{\small (a)}
\PlaceText{18mm}{60mm}{\small (b)}
\PlaceText{46mm}{60mm}{\small (c)}

We calculate the equilibrium positions $\{z_i\}$ of $N$ ions due to such potential (blue dots in Fig. 1(a)). We initialize the ion crystal at even ion separations slightly smaller than the expected $\Delta z$. We repeatedly evaluate the total force acting on each ion, and move them fractionally in that direction in order to minimize electric potential energy. The results show that the equilibrium ion separation averages to about 2.9$\mu$m with less than 5\% variation from minimum to maximum. 

Next, we find the collective transverse vibrations of the ion chain by computing the $x_i x_j$ dependence of the Hamiltonian (Fig. 1(b) and (c)). This is done by expanding electric potential and inter-ion repulsion around their equilibrium positions $\{z_i\}$, based on the Lamb-Dicke approximation ($\sqrt{\frac{\hbar}{2m\omega}}\sqrt{n+\frac{1}{2}} \ll \lambda < \Delta z$). The radial potential $V_{trap}(x) = \frac{1}{2}m\omega_x x^2$ is assumed to be constant along the z-axis, whereas the axial potential $V_{trap}(z)$ is given by equation (2). Note that $x_i z_j$ couplings vanish up to the first order, allowing us to calculate the longitudinal and transverse motional modes separately. The $x_i x_j$ terms can then be diagonalized, giving us the transverse motional modes $\hat{X}_k = \sum_{i=1}^{N}u_{ki}\hat{x}_i$ and resonant frequencies $\omega_k$ \cite{cirac1}, for $k$ from 1 to $N$.  Using the new basis $\{\hat{X}_k\}$, the total potential energy is now equivalent to a collection of $N$ non-degenerate harmonic oscillators with no phonon hopping, and the motion can be characterized by coherent displacements of these oscillators in their respective rotating frames \cite{Roos}.

We let the common mode frequency $\omega_x$ be 3.07 MHz and calculate the higher order modes using the $\{z_i\}$ previously obtained where $\Delta z \approx 2.9 \mu$m. As expected, the resonant frequencies are unevenly spaced, with the lowest frequency being 2.45 MHz. The motional modes appear to be standing waves with increasing wavenumbers, which can be explained by the periodicity of a uniform charge density. Fig. 1(b) shows the middle part of the spectrum and Figs 1(c) shows the the 25th to 27th transverse modes explicitly. 

\section{Ion motion due to a time-dependent driving force}

The M\o lmer-S\o rensen gate is performed by applying two tones with equal but opposite detunings from the carrier transition to each chosen ion \cite{Molmer, Sorensen}. For each sideband $k$, the motional state of the ion chain when the i-th ion is driven by a time-dependent external force is described as the following complex integral

\begin{equation}
\begin{aligned}
\alpha_k(t) &= \eta_{ik}\int_{0}^{t}\Omega(t')e^{i\theta_k(t')}dt', \\
\hspace{5pt} \theta_k(t) &= \int_{0}^{t}\delta_k(t')dt'
\end{aligned}
\end{equation}
where $\eta_{ik}$ is the Lamb-Dicke parameter when the i-th ion is addressed, $\Omega(t)$ is the carrier Rabi frequency for single-qubit rotation (proportional to driving intensity), and $\delta_k(t)$ is the detuning between the driving frequency $\mu(t)$ and sideband frequency $\omega_k$ \cite{Roos, my_paper}. To characterize the ions' vibrational motion, we keep track of $\alpha_k$ as a trajectory in the phase space, one for each sideband. Thus $\alpha_k(0) = 0$ and $\alpha_k(\tau)$ are the starting and end point of the trajectory respectively.

When we apply the same driving force to two ions, not only is the ion lattice set in motion, the ions also become entangled in the qubit space due to the non-commutativity of the ladder operators of motional phonons. Full entanglement can be achieved if the appropriate Rabi strength is applied. Conveniently, the degree of entanglement is proportional to the area encircled by the trajectory. However, if $\alpha_k(t)$ is non-zero at $t = \tau$ where $\tau$ is the total gate time, the qubits will remain entangled to the motional space. Without optimization, this becomes a source of gate error, which can be estimated as $\mathcal{E} = \sum_{k=1}^{N}|\alpha_k(\tau)|^2$ for $\alpha_k \ll 1$ \cite{my_paper}. 

There are many ways we can minimize $\mathcal{E}$, or $|\alpha_k(\tau)|$ for all $k$, which is a non-trivial problem for large $N$ due to the crowding of motional spectrum. In general, far-detuned modes ($\delta_k \gg 1/\tau$) will be decoupled as long as the intensity profile for $\Omega(t)$ is reasonably smooth, or if the rate of change of $\Omega$ is slower than the detuning (as opposed to step functions). Such pulses have a narrow frequency response, meaning that excitation of motional modes will vanish quickly as detuning increases. For near-detuned modes, we may minimize remaining motion by introducing a suitable number of free parameters during the gate \cite{AM1, Phase_decoupling, my_paper}. Here we propose allowing small, periodic oscillations of the applied frequency to suppress the residual motion for the 10 nearest modes.

To analyze the impact of oscillations in frequency on $\alpha_k$, we assume that the detuning pattern can be decomposed to Fourier components, expressed as follows

\begin{equation}
\begin{aligned}
\delta_k(t) &= \delta_{k,0} + \sum_{n=1}^{\infty} a_n \cos(w_n t) \hspace{12pt} \text{where} \hspace{12pt} w_n = 2n\pi/\tau \\
\end{aligned}
\end{equation}

where $a_n$ vanishes quickly with $n$ for smooth $\delta_k(t)$. Plugging equation (5) in (4), and assuming $a_n \tau \ll 1$, we may write

\begin{equation}
\begin{aligned}
\alpha_k(t)\approx \eta_{ik}\Bigg[ &\int_{0}^{t}\Omega(t')e^{i\delta_{k, 0}t'}dt' \hspace{15pt}\\
+\sum_{n=1}^{\infty} \frac{a_n}{2w_n} \Bigg(&\int_{0}^{t}\Omega(t')e^{i(\delta_{k,0}+w_n)t'}dt' \\
- &\int_{0}^{t}\Omega(t')e^{i(\delta_{k,0}-w_n)t'}dt'\Bigg)\Bigg] \\
\end{aligned}
\end{equation}

The summed terms are effectively additional tones centered around the average detuning $\delta_{k, 0}$. For large $\delta_{k, 0}$, most of these terms are negligible due to the smallness of $a_n/w_n$ and largeness of $\delta_{k, 0}\tau$ and $(\delta_{k, 0}\pm w_n)\tau$, leading to small, quickly rotating terms. This result allows us to minimize near-detuned terms by using a particular combination of $a_n$, while keeping far-detuned modes decoupled. 

\begin{figure*}

	\scalebox{0.47}{\includegraphics{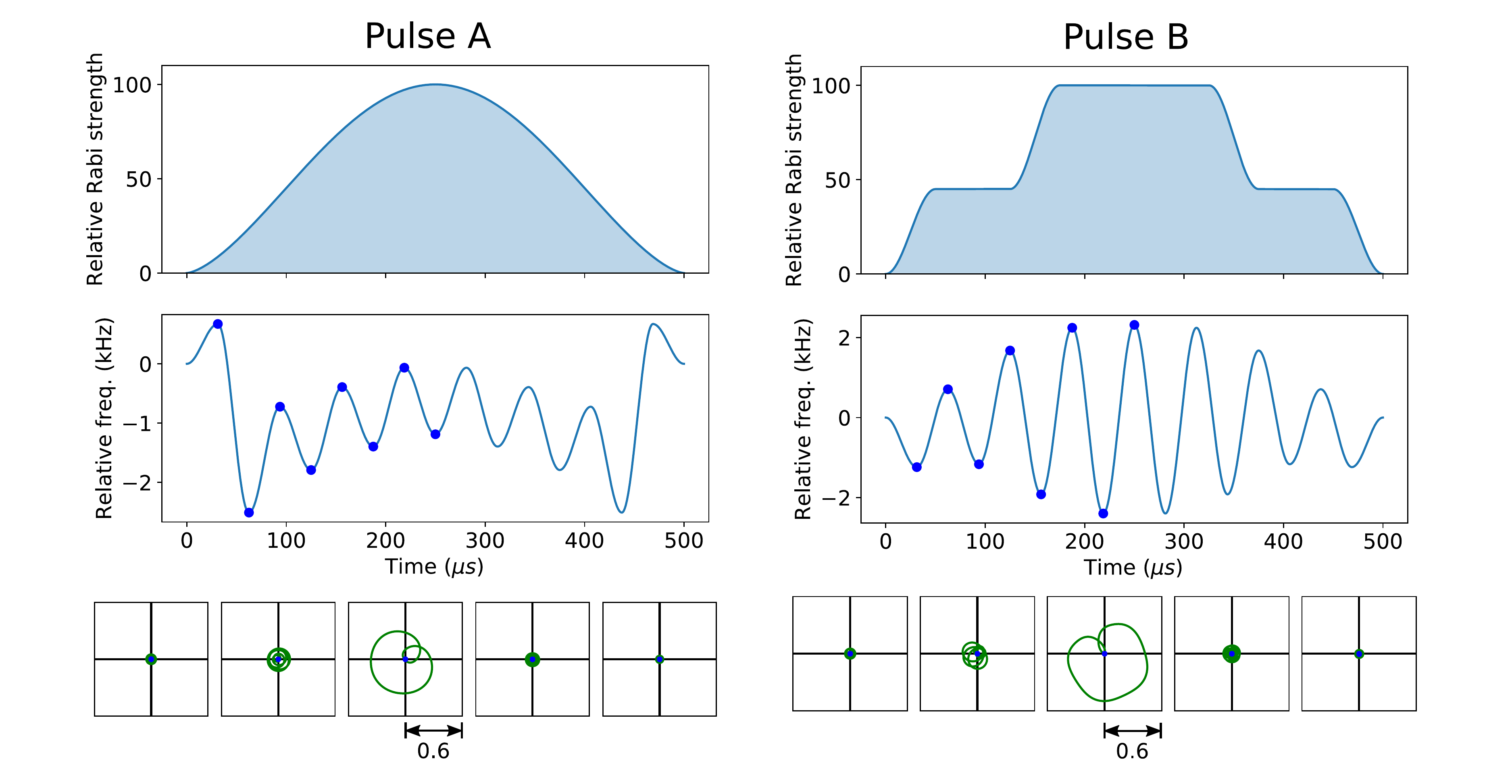}}
	\caption{\small (a) and (b) are the relative Rabi strength $\Omega(t)$ applied over a gate time of 500 us, denoted as Pulse A and B respectively. (c) and (d) are the corresponding frequency patterns ($\mu(t)-\mu_0$ where $\mu_0 = \omega_{26} - 3.7$ kHz) that minimize residual ion motion. The blue dots are allowed to move vertically during optimization. Note that the pulses are set to be symmetric in time. (e) and (f) are the resultant phase space trajectories from the 24th to 28th motional modes. Note that the detuning determines the curvature of the phase space trajectories, whereas the Rabi strength determines trajectory speed as well as curvature.}
\end{figure*}
	
\section{Optimization Procedure and Results}

In this paper, the sideband spectrum is obtained from the 50-ion simulation described in the previous section, but in the future it should be obtained by experimental measurement. This information will help us minimize gate error, but the power required for full entanglement depends on the ions chosen.

The first step is to set the shape of $\Omega(t)$. For the sake of comparison, we assume two intensity profiles, pulse A and B (Fig. 2(a) and (b)), and perform the same optimization with frequency to minimize state-dependent motion. Pulse A has a time dependence of (sin$(\pi t/\tau))^{1.5}$, whereas Pulse B consists of 3 steps connected by cosine functions. We note that Pulse A is ``smoother" than B in the sense that it has a lower maximum rate of change in intensity, which should make it more resilient against frequency offsets. In both cases,  the smoothness suppresses excitation of far-detuned modes, such that they only contribute to a small fraction of $\mathcal{E}$.

For both pulses, we need to initialize the frequency pattern and then seek a final pattern that minimizes ion motion. In this example, we choose a reference frequency $\mu_0 \approx \omega_{26} - 3.7$ kHz. We then modify the driving frequency $\mu(t)$ around $\mu_0$ periodically to minimize the displacement for the 10 nearest-detuned modes ($\alpha_{22}$ through $\alpha_{31}$). Finally, we calculate the total error due to motion in all 50 modes and confirm that $\mathcal{E} < 10^{-4}$, and test the gate's robustness against frequency offsets.

We pause here to explain the rationale behind this procedure.  The driving frequency is chosen to be near resonant with $\omega_{26}$ since all ions are excited to a similar degree in the 26th mode, making it an ideal quantum channel for multi-qubit entanglement. We note that the sideband splitting near this mode is at 18 kHz, less dense than most other parts of the spectrum, which allows us to resolve the motional modes with shorter gate times. Another important advantage of using such a high-order mode is the significantly lower heating rates due to trap noises, since typical trap electrode sizes are much larger than the average separation between neighboring ions. 

The frequency pattern is constructed as follows. We set a series of turning points (blue dots in Fig. 2(b)) at equal time intervals and connect each pair of neighboring points with cosine curves, leading to an oscillatory shape. Our algorithm adjusts the vertical positions of the extreme points until the cost function reaches a local minimum. It can be readily verified that this pattern can be described by a truncated Fourier series very accurately, so the argument using equation (6) still holds. The cost function we seek to minimize is the sum of squares of time-averaged displacements of the phase space trajectories of the 10 nearest-detuned modes. If successful the pulse will not only suppress motional displacement, but also the first-order dependence of the displacement on unwanted frequency offsets $\delta_1$ (see \cite{my_paper}). As a result, $\mathcal{E} = \sum_{k=1}^{N}|\alpha_k(\tau)|^2$ will scale as $\delta_1^4$, making the gate robust against small frequency drifts of the trap.

\PlaceText{20mm}{25mm}{\small (a)}
\PlaceText{20mm}{55mm}{\small (c)}
\PlaceText{20mm}{87mm}{\small (e)}

\PlaceText{103mm}{25mm}{\small (b)}
\PlaceText{103mm}{55mm}{\small (d)}
\PlaceText{103mm}{87mm}{\small (f)}

\begin{figure*}
	
	\scalebox{0.56}{\includegraphics{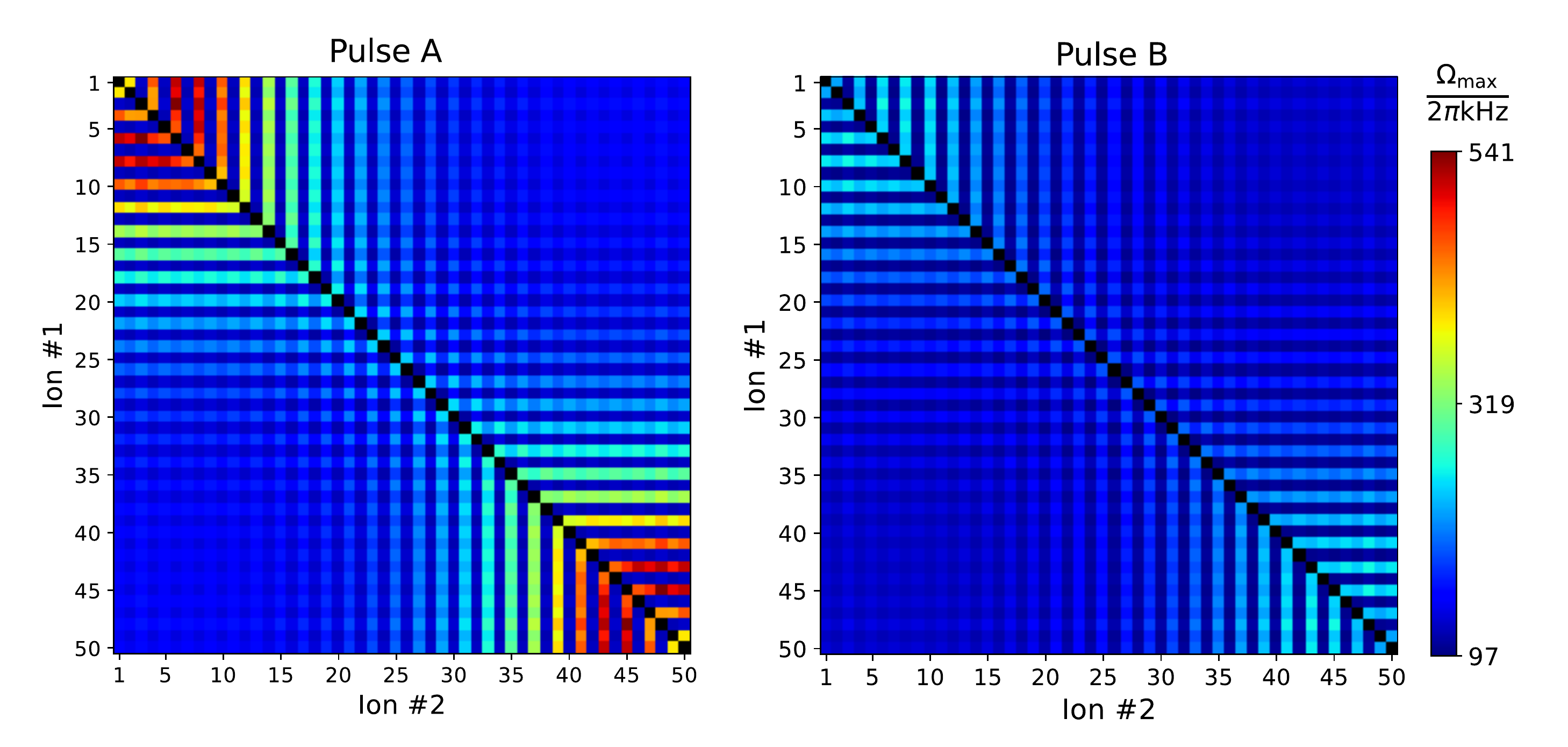}}
	\caption{\small  Rabi frequency $\Omega_{max}$ required to entangle any pair of qubits ($\Omega_{max}$ corresponds to relative Rabi strength = 100 in Fig. 2(a)). It ranges from $2\pi\times 109$ kHz to 541 kHz for Pulse A, and from $2\pi\times$ 97 kHz to 263 kHz for Pulse B.}
\end{figure*}

\begin{figure}
	\scalebox{0.5}{\includegraphics{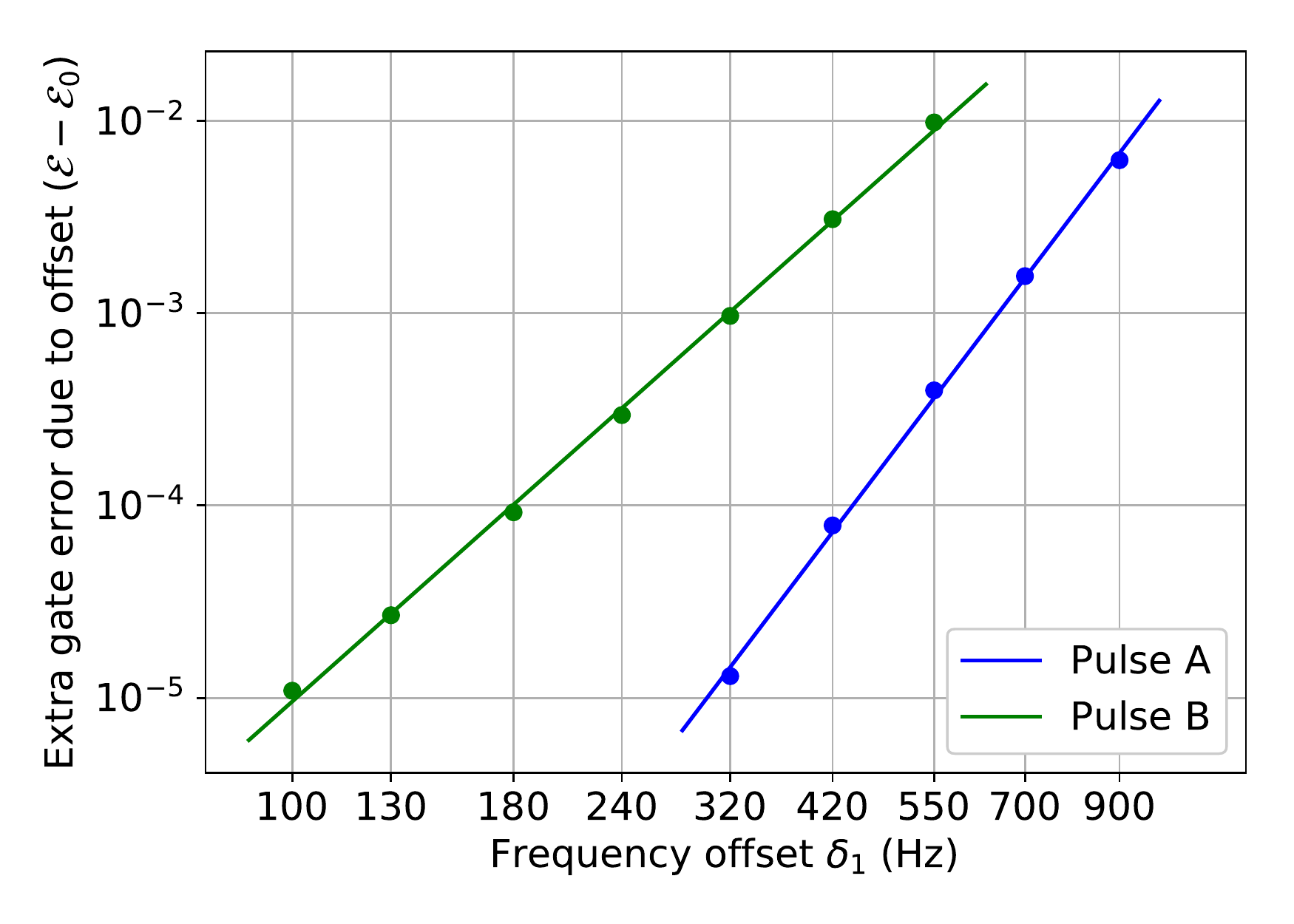}}
	\caption{\small  Log-log plot of extra gate error ($\mathcal{E} - \mathcal{E}_0$) versus frequency offset in the applied frequency ($\mu_0 \to \mu_0 + \delta_1$) for pulse A and B, where $\mathcal{E}_0$ is the gate error when there is no offset. The average slope is 5.95 and 4.01 for Pulse A and B respectively, compared to the the expected 4.}
\end{figure}

Figs. 2 (c) and (d) show the optimized frequency patterns for Pulse A and B, both consisting of 8 oscillations and thus 15 turning points. But since the pulse is set to be symmetric, there are only 8 degrees of freedom (8 blue dots) for frequency modulation. The oscillatory amplitude is about 2 kHz or less, much smaller than the average sideband splitting of 18 kHz. The initial gate error (when $\mu(t) = \mu_0$) due to residual motion is $9.4\times10^{-4}$ for Pulse A and $6.3\times10^{-3}$ for Pulse B, and the final error is $2.4\times 10^{-6}$ and $3.0\times 10^{-5}$ after optimization with $\mu(t)$. The corresponding trajectories of the 5 nearest modes are plotted in Figs. 2(e) and (f). The fact that they are centered around the origin confirms that the pulse suppresses final motion and is robust against slow frequency drifts (small change in overall curvature of trajectories).

A log-log graph is plotted in Fig. 4 of the additional gate error $\mathcal{E}-\mathcal{E}_0$ versus constant frequency offsets $\delta_1$ ($\mathcal{E} = \mathcal{E}_0$ when $\delta_1 = 0$). It is readily seen that Pulse A has a greater tolerance against frequency offsets than B, especially for lower error thresholds. The slopes found by linear regression is $5.95 \pm 0.29$ for Pulse A and $4.01 \pm 0.16$ for Pulse B, which are equal to or larger than the predicted 4. We conclude that Pulse A is more robust against frequency errors than Pulse B, as expected.

The colored graph in Fig. 4 shows the power or maximum Rabi strength required to entangle any pair of ions in the chain ($\Omega_{max}$ such that $\beta_{ij} = \pi/4$). All pairs can be entangled with power $\leq 2\pi$ $\times$ 541 kHz for Pulse A, and $\leq 2\pi$ $\times$ 263 kHz for Pulse B, implying that the latter pulse has higher coupling efficiency. This is partly due to a higher overall detuning of the optimized frequency pattern from $\omega_{26}$ for Pulse A than for Pulse B (see Figs. 2(c) and (d)), which leads to a greater enclosed area by the phase space trajectory for Pulse B (Figs. 2(e) and (f)). The required power does not increase as a function of distance. Instead, it alternates between low and high, and averages to roughly $2\pi$ $\times$ 150 kHz for long distances. The required power is also higher towards the edges of the ion chain. To summarize, for the same gate time and degrees of freedom, Pulse A consumes less power, but Pulse B has higher tolerance against frequency errors. This flexibility with the initial conditions allows trade-off between robustness and power efficiency.

\section{Conclusion}

We have provided a suite of tools that can predict the distribution and motion of a 1-D ion lattice and search for pulses that entangle any pair of ions with high fidelity and reasonable overhead. Despite the crowding of resonant motional mode spectrum, the interaction strength between an ion pair does not decrease with distance, and maximal entanglement is achieved with finite driving power. The residual motion of the lattice can be suppressed efficiently by using an optimized pulse with modulated amplitude and frequency. At least two distinct solutions have been found with limited degrees of freedom, showing the balance between gate power scaling and robustness against frequency errors.

Unwanted frequency offset is only one of the many error sources we observe with ion traps, which also include intensity fluctuations and trap heating. These can be treated by deploying similar physical control methods \cite{robust_gates, resilient_gates, hf_heating}. We can also mitigate such effects with other control sequences such as Walsh modulation \cite{coherent_suppression} or two-qubit composite pulse sequences \cite{robust_ising, yu1, comp_2_qubit}.

It is also important to note that the motional mode structures are sensitive to trap imperfections, and we argue that our method does not lose generality because of this. For example, we may not be able to generate the potential in equation (2) which leads to uniform ion density with arbitrary accuracy. Also, the radial confinement may also not be perfectly uniform across the length of the ions. These deviations will considerably alter the mode structures, meaning that ion participation will not be as predictable as shown in Fig. 1(c). This dilemma can be solved by searching for alternative pulses with different starting frequencies $\mu_0$, so that any pair of ions can be entangled through at least one of the pulses, since each ion is more involved in some sidebands than the others. 

As we advance towards 100 ions or more, there are many proposals for dividing ions into groups and establish entanglement between them. Instead of further enlarging the trap, we may rely on modular approaches such as zoning and shuttling of ions \cite{shuttle1, shuttle2, large_scale, fast_transport, coherent_diabatic} as well as photonic links \cite{photon_links, freely_scalable}. Whichever direction we go, the ability to entangle ions arbitrarily within the same trap will vastly improve the scalability of ion traps as a quantum computer.

\section{Acknowledgements}
The authors thank Kevin Landsman, Norbert Linke, Luming Duan, and Ivan Deutsch for useful discussions. This work was supported by the Office of the Director of National Intelligence - Intelligence Advanced Research Projects Activity through ARO contract W911NF-10-1-0231 and the National Science Foundation Expeditions in Computing award 1730104.

\end{document}